\documentclass[
  aip,
  amsmath,amssymb,
  preprint,
  reprint
]{revtex4-1}

\usepackage{graphicx}
\usepackage{dcolumn}
\usepackage{bm}
\usepackage[utf8]{inputenc}
\usepackage[T1]{fontenc}
\usepackage{mathptmx}
\usepackage{etoolbox}
\usepackage[colorlinks=true, allcolors=blue]{hyperref}
\usepackage{tikz}
\usetikzlibrary{arrows.meta,positioning,shapes.geometric,fit}
\usepackage[caption=false]{subfig} 
\usepackage{multirow}
\usepackage{natbib} 

\begin{document}

\preprint{AIP/123-QED}

\title[Sample title]{Multifractal features of multimodal cardiac signals: Nonlinear dynamics of exercise recovery}

\author{A.~Maluckov}
\email{sandram@vin.bg.ac.rs}       
\author{D.~B.~Stojanovi\'c}
\author{M.~Mileti\'c}
\author{M.~D.~Ivanovi\'c}
\author{Lj.~Had\v{z}ievski}
\author{J.~Petrovi\'c}
\email{jovanap@vin.bg.ac.rs}      

\affiliation{
Vin\v{c}a Institute of Nuclear Sciences – National Institute of the Republic of Serbia,
University of Belgrade, Belgrade, Serbia
}

\date{\today}
\begin{abstract}
We investigate the recovery dynamics of healthy cardiac activity after physical exertion using multimodal biosignals recorded with a polycardiograph.
Multifractal features derived from the singularity spectrum capture the scale-invariant properties of cardiovascular regulation.
Five supervised classification algorithms—Logistic Regression (LogReg), Support Vector Machine with RBF kernel (SVM-RBF), $k$-Nearest Neighbours (kNN), Decision Tree (DT), and Random Forest (RF)—were evaluated to distinguish recovery states in a small, imbalanced dataset.
Our results show that multifractal analysis, combined with multimodal sensing, yields reliable features for characterizing recovery and points toward nonlinear diagnostic methods for heart conditions.
\end{abstract}
\keywords{
Cardiac dynamics;
Nonlinear systems;
Singularity spectrum;
Multimodal biosignals;
Learning algorithms;
Recovery assessment;
Complex physiology
}

\maketitle

\begin{quotation}
\noindent

The human heart exhibits complex nonlinear dynamics, 
emerging from coupled physiological and psychological regulation mechanisms. 
Cardiac variability reflects a subtle balance between deterministic feedback loops 
and stochastic fluctuations, a signature of healthy cardiovascular function. 
Understanding how this complexity changes during recovery from physical activity
provides valuable insight into the adaptive capacity of the autonomic nervous system 
and early signatures of pathological states. 

Here, we investigate the cardiac recovery of healthy subjects using multimodal biosignals recorded with a polycardiograph, including electrocardiogram (ECG), photoplethysmogram (PPG), seismocardiogram (SCG), and phonocardiogram (PCG).
We apply multifractal analysis based on the singularity spectrum to quantify 
scale-invariant fluctuations in cardiac dynamics and assess their evolution during recovery. 
To relate the intrinsic structure of the nonlinear features to the diagnostically relevant observables, classification based on supervised learning is employed. Due to the small size and imbalance of the available dataset, five learning algorithms are tested. 
Our findings demonstrate that multifractal descriptors from multimodal signals 
capture nonlinear properties of cardiac regulation, providing 
a framework for future diagnostic tools for pathological heart dynamics.

\end{quotation}

\section{Introduction}

The complexity rooted in Nature's simplicity manifests profoundly in human physiology, 
carrying embedded information about the dynamical, physical and psychological state of the host.  Self-similarity, universal scaling laws, irregular fluctuations, 
and the coexistence of stochasticity with deterministic patterns 
are observed across physical systems and biological processes 
\cite{Bak1987SelfOrganized, Mandelbrot1982FractalGeometry, Stanley1999Scaling}.
How to extract and interpret this information to ensure stable and healthy life conditions 
remains one of the key challenges humanity faces. 
Beyond its fundamental scientific implications, this knowledge holds direct importance for medical diagnostics, early detection of dysfunction, and the design of interventions that support long-term cardiovascular and systemic health.

Methods of nonlinear dynamics, chaos theory, and advanced signal analysis in biomedical research
became essential tools for discovering the underlying mechanisms of cardiovascular and autonomic regulation
\cite{Goldberger1990Fractals, Ivanov1999Multifractality, Peng1995Mosaic, Costa2002MultiscaleEntropy}. Features such as entropy measures, fractal and multifractal descriptors, and recurrence-based indices are extracted from multimodal biosignals to quantify the dynamic structure of physiological states and transitions.

Recovery from physical activity offers a particularly valuable natural experiment for probing cardiac dynamics \cite{cole1999heart,Malik1996HRV,Shaffer2017HRV,Ivanov1999Multifractality,Goldberger1990Fractals,Mandelbrot1982FractalGeometry,Stanley1999Scaling}.
It involves the coordinated re-synchronization of electrical, mechanical, and vascular subsystems, modulated by autonomic feedback and influenced by metabolic and musculoskeletal processes. The evolving complexity of cardiac dynamics during this phase reflects the system’s adaptive capacity—and may provide early signatures of dysregulation before clinical symptoms emerge.

In this study, we investigate the cardiovascular state of healthy subjects
using a polycardiograph device equipped with four synchronized sensors:
electrocardiography (ECG), photoplethysmography (PPG), seismocardiography (SCG), and phonocardiography (PCG).
This synchronous multimodal acquisition provides complementary views of cardiac dynamics,
offering a richer description of underlying physiological processes.
Our work represents a preparatory step toward enhanced cardiovascular diagnostics and long-term health assessment, laying the basement for analyses of pathological states and their early signatures in recovery dynamics.

Beyond conventional the heart rate variability (HRV) measures 
\cite{Malik1996HRV, Shaffer2017HRV}, 
our approach adopts the nonlinear, multiscale nature of cardiac dynamics. 
We employ multifractal analysis, which is utilized to describe 
complex behavior in natural phenomena, from fluid turbulence 
\cite{Frisch1995Turbulence, Muzy1994MF} 
to neural networks \cite{Sporns2004ComplexBrain}, 
to probe the dynamical complexity of cardiac regulation before and after exercise. 

Section~II presents the mathematical tools and methods utilized in this study, 
enabling the extraction of relevant features from the multifractal (singularity) spectra. By ranking these features and their combinations we prepare the conditions for classification of recovery conditions which is necessary due to the small and unbalanced dataset. Section III presents the results of the singularity spectrum analysis 
of recorded signals from all sensors and relate them to the physiological state of the subjects. 
Section IV discusses the classification performances of different learning algorithms. 
Finally, Section V concludes the study.

\section{Model Formulation and Methods}
\label{sec:methods}

In this section we present the experimental basis of our analysis, the 
multifractal formalism, the set of features extracted from the singularity spectrum, 
and the classification framework (Fig. \ref{fig:workflow}). The aim is to provide a self–contained and 
reproducible formulation of the methods applied to multimodal cardiovascular signals.

\subsubsection{SensSmartTech database}

We analyze 301 recordings of 
30~s each from 28 healthy volunteers under resting and post–
exercise conditions from the SensSmartTech database \cite{senssmarttech_physionet, LazovicSciData2025}. The signals were acquired with a 
polycardiograph, a device that synchronously records ECG, PPG, SCG, 
and PCG at sensor-specific sampling rates (ECG/SCG 500~Hz, 
PPG 100~Hz, PCG 1000~Hz).  

ECG signals were recorded from a subset electrodes including four limb and V3, V4 precordial electrodes. The remaining four precordial electrodes were not used and were placed on the upper arm to avoid excessive noise. PCG was placed near the xiphoid extension and supported by an elastic tape. SCG was recorded by an accelerometer mounted by an adhesive snap ECG electrode between the V3 and V4 ECG electrodes, near the heart apex. PPG signals were acquired at the carotid (PPGc) and brachial artery (PPGb), enabling 
cross–site assessment of pulse–wave characteristics. The PPGs provided recordings taken at 660 nm and 880 nm. In this work, we selected the 660 nm channel due to a higher signal-to-noise ratio computed fol-
lowing the methodology described in~\cite{LazovicSciData2025, petrovic2024validation}.

All signals were preprocessed prior to analysis. The ECG 
was denoised using the iterative regeneration method~\cite{petrovic2024validation} 
and high--pass filtered at 0.5~Hz to suppress baseline wander. 
PPG signals were band--pass filtered in the range 0.5--6~Hz, with additional 
time--average smoothing (short moving--average window) applied to PPG\textsubscript{b}. 
SCG was band--pass filtered in the range 0.5--25~Hz, and PCG in the range 
20--150~Hz to remove baseline drift and high--frequency noise. 
All filters were 4th--order IIR Butterworth filters implemented in zero--phase form 
using MATLAB. Figure~\ref{fig:signals} presents a representative set of preprocessed 
signals for one subject.

\subsubsection{Multifractal Formalism and Singularity Spectrum}

To quantify the nonlinear and scale-invariant properties of the physiological signals, we applied the multifractal formalism based on wavelet-leader scaling analysis 
\cite{kantelhardt2002multifractal,Ivanov1999Multifractality}. 
Given a discrete time series \(x(t)\), we first computed the partition function
for a set of moments \(q \in \mathbb{R}\) across scales \(a\):
\begin{equation}
    Z(q, a) = \sum_{i} \mu_i(a)^q,
    \label{eq:partition}
\end{equation}
where \(\mu_i(a)\) denotes the local measure of fluctuations (e.g., wavelet coefficients) 
at scale \(a\) and position \(i\). 
The scaling behavior of \(Z(q,a)\) defines the multifractal scaling exponent \(\tau(q)\):
\begin{equation}
    Z(q, a) \propto a^{\tau(q)} \quad \text{for} \quad a \to 0.
    \label{eq:tau}
\end{equation}

The multifractal singularity spectrum $D(h)$—also called the Hölder spectrum—quantifies 
the distribution of local regularities in a signal \cite{Lichtenberg1992} . It is obtained from the scaling exponent 
function \(\tau(q)\) via a Legendre transform:
\begin{align}
    h(q) &= \frac{d\tau(q)}{dq}, \label{eq:holder}\\
    D(h(q)) &= q\,h(q) - \tau(q), \label{eq:spectrum}
\end{align}
where \(h(q)\) is the local Hölder exponent, a measure of the pointwise smoothness 
or singularity strength of the signal. Subsets of the signal that share the same 
\(h\) correspond to regions of similar local regularity, and \(D(h)\) gives the fractal 
dimension of each such subset. A broad \(D(h)\) indicates a wide variety of local 
dynamics, while a narrow \(D(h)\) suggests nearly uniform regularity.

\subsubsection{Multifractal Features}

From $D(h)$ (Fig. 3), we extracted several 
features characterizing the complexity and variability of cardiovascular dynamics:

\begin{itemize}
    \item Position of the spectrum maximum:
    \begin{equation}
        h_{\text{max}} = \arg\max D(h),
        \label{eq:hmax}
    \end{equation}
    which reflects the most dominant local scaling exponent in the signal.
    
    \item Spectrum width:
    \begin{equation}
        W = h_{\text{right}} - h_{\text{left}},
        \label{eq:width}
    \end{equation}
    where \(h_{\text{left}}\) and \(h_{\text{right}}\) are the Hölder exponents 
    at which \(D(h)\) achieves the minimum on the left and right sides of the maximum, respectively, 
    quantifying the range of multifractality.

    \item Left and right half widths: 
    $$
        \text{LHW} = h_{\text{max}} - h_{\text{left}}, 
        \text{RHW} = h_{\text{right}} - h_{\text{max}},
      $$ 
    capturing the short and long-scale fluctuations.

    \item Asymmetry:
    \begin{equation}
        A = \frac{1}{\sigma^3} \sum_{i} \big(h_i - \mu\big)^3 \, D(h_i),
        \label{eq:asymmetry}
    \end{equation}
    where $\mu$ and $\sigma$ denote the mean and standard deviation of $h$, weighted by $D(h)$. 
    $A$ quantifies skewness toward 
    strong (left-sided) or weak (right-sided) fluctuations.
    
    \item Area under the spectrum (AUC):
    \begin{equation}
        \text{AUC} = \int_{h_{\text{left}}}^{h_{\text{right}}} D(h) \, dh,
        \label{eq:auc}
    \end{equation}
    representing the overall fractal dimension encompassed by the spectrum.
\end{itemize}

Such scaling markers have long been linked to cardiovascular complexity and autonomic regulation \cite{Ivanov1999Multifractality,Goldberger1990Fractals}. 

Among the commonly extracted measures, triplet ($h_{\max}$,  $W$,  $A$)  captures nearly orthogonal aspects of the spectrum: location of the dominant Hölder exponent, overall heterogeneity of scaling exponents, and imbalance between short- and long-scaled fluctuations, respectively. This motivates our focus on $h_{max}, W, A$ as the most informative set of multifractal features.

\subsection*{ Multifractality vs. Nonlinear Dynamics}

To check whether multifractality in multimodal biosignals reflects nonlinear dynamics rather than statistical artifacts, we compared multifractal 
descriptors with recurrence‐based measures of nonlinear structure. 

For each subject and modality (ECG, PPGc, PPGb, SCG, PCG), every recovery 
signal was processed to extract the core multifractal descriptors 
($h_{\max}$, $W$, $A$, and  AUC) using 
wavelet‐leader multifractal analysis.  
From the same signals, recurrence quantification analysis (RQA) indices 
(determinism DET, laminarity LAM, recurrence rate RR, entropy ENTR, and 
mean diagonal line length $L$) were computed after phase‐space reconstruction.  

Both descriptors were obtained on a one‐to‐one basis for every recording, 
and their correlations were assessed at subject‐ and modality‐levels.  
To further verify that multifractality was not an artifact of linear correlations 
or amplitude distributions, we performed a surrogate analysis 
(IAAFT surrogates and shuffled surrogates) \cite{Schreiber1996,Theiler1992}, 
comparing multifractal descriptors from the original signals with those from 
surrogates preserving the power spectrum but destroying nonlinear structure.

\subsubsection{Classification Framework}

\begin{figure}[ht!] 
  \centering
 \includegraphics[width=9cm]{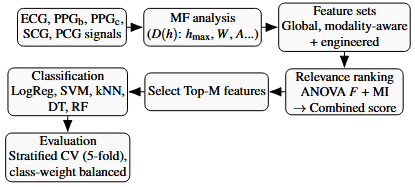}
  \caption{Classification scheme.}
  \label{fig:workflow}
\end{figure}

To map multifractal features to recovery state, we first define heart rate (HR) 
as the number of cardiac cycles per minute, computed from successive RR intervals 
in the ECG:
\begin{equation}
    \text{HR} = \frac{60}{<RR>},
\end{equation}
where \(<RR>\) is the average time (in seconds) between consecutive R peaks in the ECG signal.
Heart-rate recovery (HRR) thresholds~\cite{cole1999heart} were then used as 
reference categories:
\begin{align*}
    \text{GOOD} &: \text{HRR} > 25 \; \text{bpm}, \\
    \text{AVERAGE} &: 13 \leq \text{HRR} \leq 25 \; \text{bpm}, \\
    \text{LOW} &: \text{HRR} < 13 \; \text{bpm}.
\end{align*}

The HRR for each ECG recording was quantified as
\begin{equation}
    \text{HRR} = \text{HR}_{\text{max}} - \text{HR}_{\text{after 1 min}},
    \label{eq:hrr}
\end{equation}
where $\text{HR}_{\text{max}}$ denotes the peak heart rate just after running and 
$\text{HR}_{\text{after 1 min}}$ the heart rate measured one minute after. 
To ensure stable estimates, post-exercise HR sequences from the first ECG lead 
were smoothed using a low-order polynomial fit \cite{cole1999heart,Shaffer2017HRV}.

Our goal is to extract information on cardiovascular recovery and adaptation 
after exercise. Since recovery is a dynamical process involving autonomic and 
hemodynamic regulation, the features must capture different aspects of this 
temporal evolution. We define two complementary feature sets:
\begin{enumerate}
  \item Net–change descriptors $F_{\text{change}}$, defined as the difference between 
  the first and last post–exercise value of each multifractal feature.
  \item Half–recovery descriptors $F_{\text{half}}$, derived from fitting exponential 
  or polynomial recovery curves and evaluating each subject at the median cohort 
  half–recovery time.
\end{enumerate}

In this study, feature relevance was assessed with filter methods combining the analysis of variance (ANOVA) 
$F$-statistics and mutual 
information (MI) into a unified 
score. This procedure identifies features with both linear discriminative power and nonlinear associations, while avoiding overfitting in the small datasets.

To evaluate the discriminative power of the multifractal dynamical features ($F_{\text{change}}$, $F_{\text{half}}$),  we applied three feature ranking and selection strategies prior to classification: global ranking of all multifractal features across modalities, per-modality ranking and ranking over the extended multimodal feature space. In the last, the engineered features are determined as simple nonlinear transformations (e.g., squared terms) and cross-modality combinations (products) of original features. The top $M$ ranked features were then selected for classification in each strategy. 
These strategies were designed to contrast global feature dominance with modality-specific balance and to test whether simple nonlinear interactions could enhance classification performance,  (Fig. \ref{fig:workflow}). 

Physiologically, these categories map multifractal features and HRR into indicators of the heart’s adaptive capacity, bridging signal dynamics with clinically meaningful fitness states~\cite{cole1999heart,Goldberger1990Fractals}.

\subsubsection{Classification Algorithms and Numerical Implementation}

Classification was performed using five supervised algorithms: Logistic Regression, 
Support Vector Machine with an RBF kernel, $k$--Nearest Neighbours, Decision Tree, 
and Random Forest~\cite{Bishop2006,Hastie2009} (Fig.~\ref{fig:workflow}). 
The purpose of this study was exploratory---to assess whether recovery 
dynamics can differentiate fitness states. Given the limited dataset size, 
independent training and test sets were not created. Instead, model performance 
was evaluated with stratified 5--fold cross--validation, reporting accuracy, 
balanced accuracy, and macro--F1%
~\cite{Bishop2006,Hastie2009,Powers2011,Pedregosa2011}. 
Macro--F1, defined as the harmonic mean of precision and recall averaged across 
classes, is particularly suited for imbalanced datasets since it remains 
sensitive to errors in minority categories~\cite{Powers2011}. 
Class--weight balancing was applied so that minority categories 
(e.g., \textit{LOW}) contributed proportionally more to the loss 
($\approx$3$\times$ relative to \textit{AVERAGE}), which improved macro--F1 
stability and reduced neglect of rare classes.

Multifractal spectra were computed using the DWTLEADER algorithm with 
a biorthogonal spline wavelet filter (bior1.5) in Matlab, chosen for its 
good time--frequency localization and near-symmetry. 
Structure functions were estimated for $q \in [-5,5]$, with decomposition 
restricted to levels providing at least six wavelet leaders. 
Feature extraction and classification were implemented in Python with 
\texttt{scikit-learn}~\cite{Pedregosa2011}.

\section{Results}

\subsection{Dynamics of the Singularity Spectrum During Recovery}

Figure~\ref{fig:spectrum} illustrates the multifractal singularity spectra $D(h)$ 
for one representative subject across all modalities. 
The pre-exercise baseline is shown in blue, spectra from early recovery 
in black, and spectra from later recovery in red.  

Across all modalities, exercise induced systematic changes 
in the multifractal descriptors. Immediately after running 
(black curves in Fig.~\ref{fig:spectrum}), the dominant singularity exponent 
$h_{\max}$ shifted leftward, reflecting more irregular dynamics. 
Concurrently, the spectrum width $W$ broadened, 
indicating greater heterogeneity of fluctuations, while the asymmetry $A$ 
skewed toward stronger (left-sided) variations. 
As recovery progressed (red curves), $h_{\max}$ gradually increased toward baseline, 
$W$ narrowed, and $A$ moved back toward symmetry. 
These trends were most pronounced in ECG and carotid PPG, 
moderate in brachial PPG, and weaker but still detectable in SCG and PCG.  

Taken together, the trajectories of $h_{\max}$, $W$, and $A$ 
provide complementary markers of cardiovascular recovery: 
$h_{\max}$ captures the degree of regularity, 
$W$ reflects the diversity of fluctuations, 
and $A$ quantifies their balance across scales.

\begin{figure}[ht!]
\centering
\includegraphics[width=9cm]{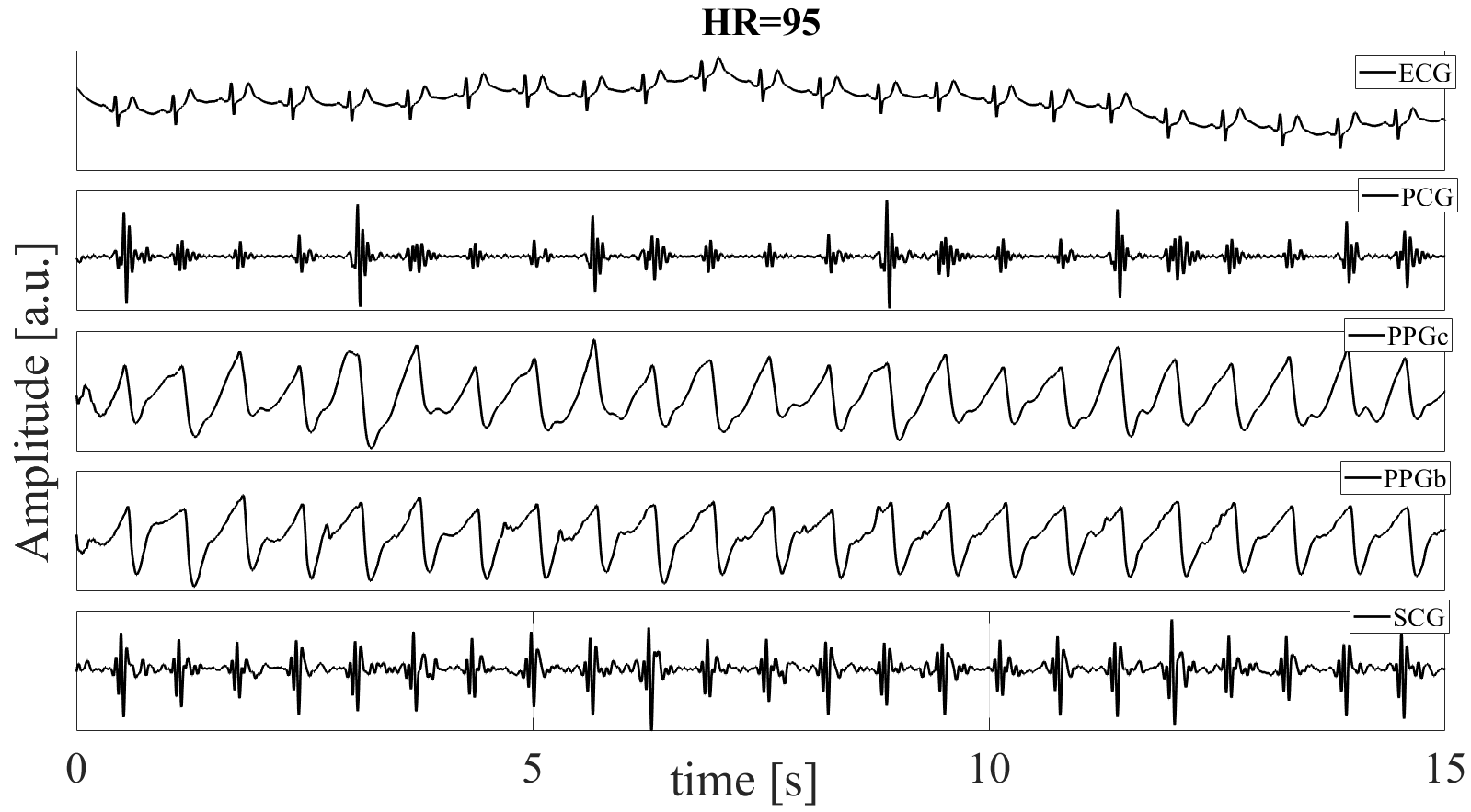}
\caption{Example of part of $30$-second raw post-running signals recorded simultaneously by the polycardiograph: ECG, PCG, PPGc, PPGb and SCG.}
\label{fig:signals}
\end{figure}

\begin{figure}[ht!]
\centering
\includegraphics[width=0.8\linewidth]
{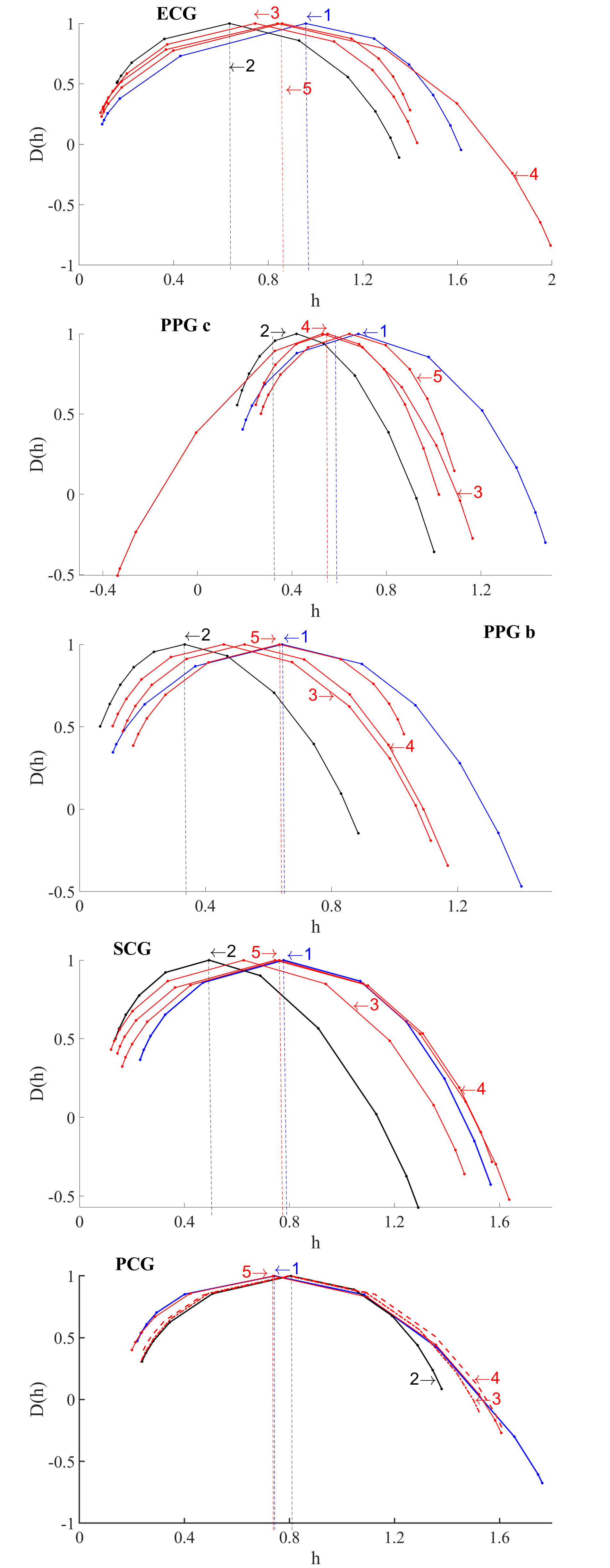} 
\caption{Multifractal singularity spectra $D(h)$ for ECG, PPGc, PPGb, SCG, 
and PCG signals of one representative subject during recovery after exercise. 
The pre-exercise signal spectra is shown in blue, spectra from early recovery 
in black, and spectra from later recovery in red. 
Vertical dashed lines denote the positions of $h_{\max}$ for each signal. 
The sequence of curves reflects the temporal order of recovery.}
\label{fig:spectrum}
\end{figure}
\subsection{Evidence: Nonlinearity Shapes 
Multifractality}

Using the full dataset of 30\,s post‐exercise recordings across all modalities (ECG, PPGc, PPGb, SCG, and PCG) and subjects, we compared multifractal descriptors ($h_{\max}$, $W$, $A$, AUC) with recurrence quantification analysis (RQA) indices (DET, LAM, RR, ENTR, $L$).  
The MF and RQA features were computed on a one‐to‐one basis for each signal, ensuring strict pairing between scaling properties and nonlinear recurrence structure.

\begin{figure}[ht!]
\centering
\includegraphics[width=1\linewidth]{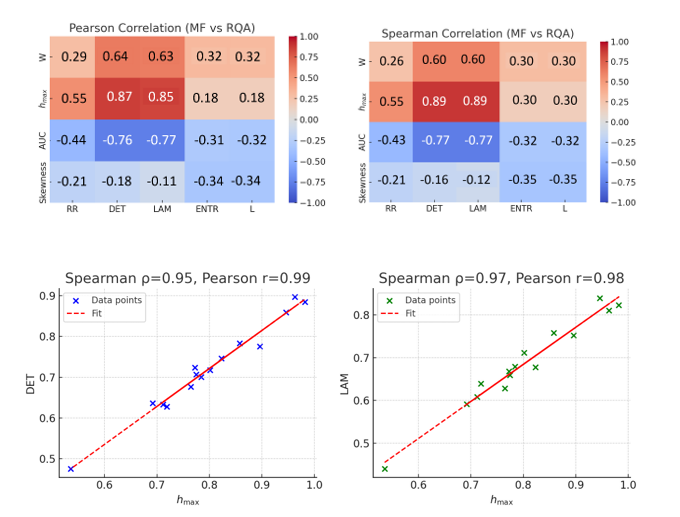}
\caption{Example of MF–RQA relationships for ECG signals of a single subject. 
Heatmaps show Spearman (left) and Pearson (right) correlations between MF and 
RQA metrics; scatter plots highlight the strongest associations ($h_{\max}$ vs.\ DET, LAM).}
\label{fig:mf_rqa_example}
\end{figure}

Across the cohort, correlations varied between modalities and subjects but revealed consistent tendencies.  
Moderate associations between $h_{\max}$ and recurrence measures such as DET and LAM were frequent (Spearman $\rho$ values often between $0.4$–$0.7$).  
Width $W$ and asymmetry $A$ showed weaker yet non‐negligible links to DET and RR, while AUC rarely exceeded surrogate‐based expectations.  
To illustrate the range of observed behavior, Fig.~\ref{fig:mf_rqa_example} shows an example from one subject’s ECG signals.

Surrogate testing \cite{Theiler1992,Schreiber1996} (IAAFT and shuffled surrogates) provided a complementary check: in a substantial fraction of recordings, the original multifractal $W$ and $h_{\max}$ values exceeded those of their surrogates at the 95\% level ($p<0.05$).  
This indicates that the observed multifractality is not simply a consequence of linear correlations or amplitude distributions. but reflects nonlinear temporal organization.  
However, the effect was not uniform: some subjects or modalities exhibited surrogate‐level behavior, underscoring variability in physiological responses.

Together, these results suggest that multifractality in post‐exercise cardiovascular signals is, at least in part, shaped by underlying nonlinear dynamics.  
Given inter‐subject variability and the limited sample size, we present these findings as supporting evidence rather than a principal conclusion, providing motivation for future studies with larger and more homogeneous cohorts.

\subsection{Classification with $F_{\text{change}}$ Features}

The $F_{\text{change}}$ descriptors represent net differences in 
multifractal parameters between the beginning and end of recovery. 
This simple definition captures the overall adaptation process but 
does not include information about intermediate dynamics.  

As shown in Table~\ref{tableI}, 
classification using raw $F_{\text{change}}$ features—either globally top-ranked (S1) or top-ranked per modality (S2)—produced only modest 
results (macro–F1 typically below $0.45$). However, when engineered features (Table~\ref{tableI}, S3),
such as nonlinear terms and cross-modality ratios, were added, performance 
improved substantially. Random Forests, in particular, reached macro–F1 
values above $0.60$, while SVM-RBF and LogReg also benefitted.  

The confusion matrix for the best-performing model (Random Forest, Fig.~\ref{fig:cm-dt-rf}) 
further illustrates this improvement. The classifier correctly identified 
$14$ out of $16$ \texttt{AVERAGE} recovery cases and showed reasonable separation 
between \texttt{GOOD} and \texttt{LOW} categories. 
Nonetheless, some confusion between \texttt{GOOD} and \texttt{AVERAGE} labels remained, 
suggesting overlap in their $F_{\text{change}}$ patterns. This likely reflects 
individual variability in recovery dynamics, such as differences in baseline fitness 
or autonomic responsiveness.

These results highlight that while raw $F_{\text{change}}$ descriptors 
offer limited discrimination, engineered extensions reveal nonlinear 
dependencies across modalities that are highly informative 
for distinguishing recovery states.

\begin{figure}[ht!]
\centering
\includegraphics[width=1\linewidth]{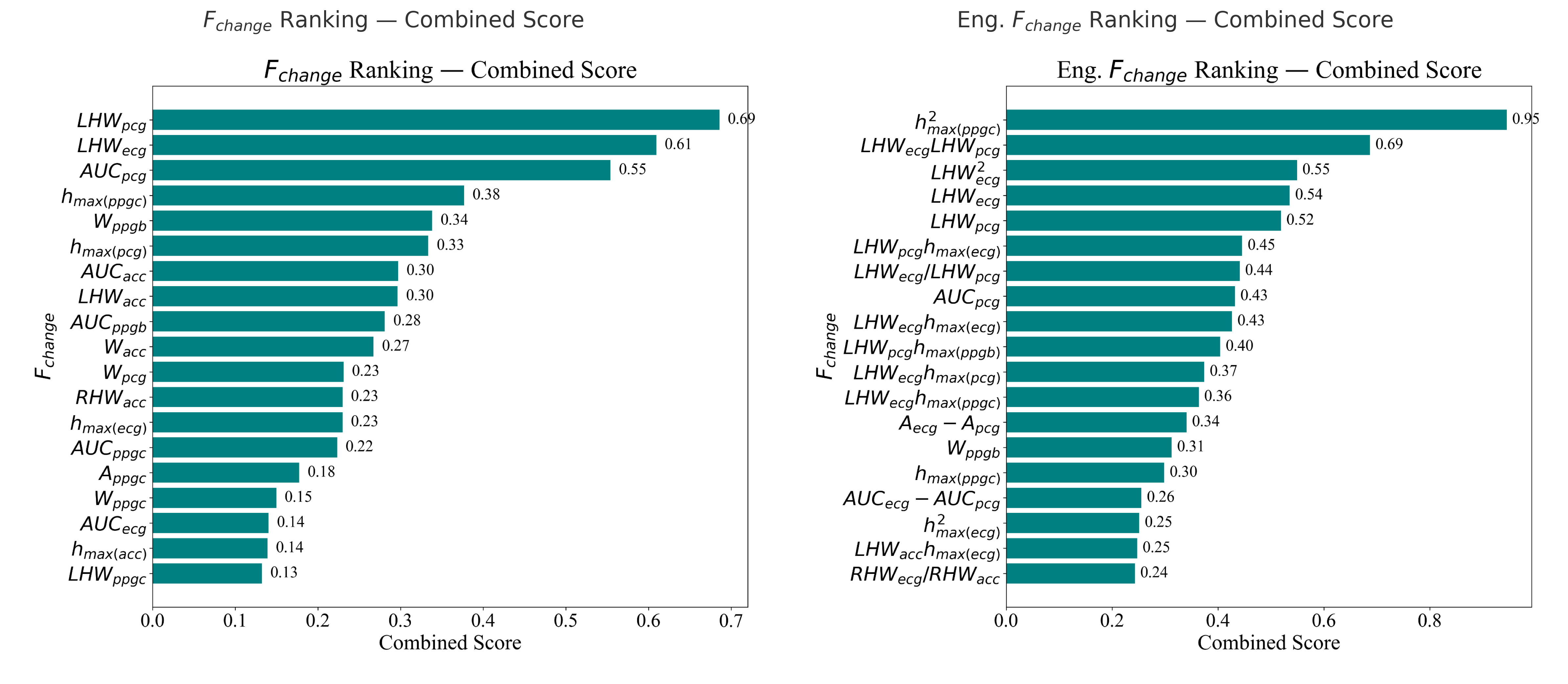}
\caption{Comparison of original vs engineered $F_{change}$ features ranked by Combined Score.}
\label{fig:feat-ranking-dp}
\end{figure}

\begin{table*}[htbp]
\centering
\resizebox{\textwidth}{!}{
\begin{tabular}{l|ccc|ccc|ccc}
\hline
\multirow{2}{*}{Model} & 
\multicolumn{3}{c|}{Accuracy} & 
\multicolumn{3}{c|}{Macro-F1} & 
\multicolumn{3}{c}{Balanced Acc.} \\
 & S1 & S2 & S3 & S1 & S2 & S3 & S1 & S2 & S3 \\
\hline
RF      & 0.67$\pm$0.18 & 0.60$\pm$0.23 & 0.74$\pm$0.20 & 
          0.46$\pm$0.20 & 0.43$\pm$0.22 & 0.62$\pm$0.28 &
          0.51$\pm$0.19 & 0.47$\pm$0.21 & 0.66$\pm$0.28 \\
kNN     & 0.60$\pm$0.21 & 0.59$\pm$0.24 & 0.63$\pm$0.14 &
          0.37$\pm$0.21 & 0.39$\pm$0.20 & 0.44$\pm$0.14 &
          0.42$\pm$0.20 & 0.43$\pm$0.21 & 0.49$\pm$0.15 \\
LogReg  & 0.45$\pm$0.19 & 0.55$\pm$0.22 & 0.50$\pm$0.12 &
          0.36$\pm$0.21 & 0.47$\pm$0.22 & 0.40$\pm$0.13 &
          0.40$\pm$0.26 & 0.51$\pm$0.22 & 0.43$\pm$0.14 \\
SVM-RBF & 0.42$\pm$0.12 & 0.48$\pm$0.23 & 0.39$\pm$0.16 &
          0.33$\pm$0.15 & 0.38$\pm$0.20 & 0.34$\pm$0.11 &
          0.33$\pm$0.14 & 0.42$\pm$0.18 & 0.35$\pm$0.13 \\
DT      & 0.35$\pm$0.09 & 0.52$\pm$0.22 & 0.53$\pm$0.14 &
          0.28$\pm$0.09 & 0.45$\pm$0.21 & 0.51$\pm$0.09 &
          0.33$\pm$0.12 & 0.52$\pm$0.15 & 0.57$\pm$0.11 \\
\hline
\end{tabular}
}
\caption{Model performance comparison under three $F_{\text{change}}$ feature-based strategies: 
global top 5 (S1), per-modality top 5 (S2), engineered top 3 (S3). 
Values are mean $\pm$ SD across 5-fold CV.}
\label{tableI}
\end{table*}

\begin{figure}[htbp]
\centering
\includegraphics[width=0.5\textwidth]{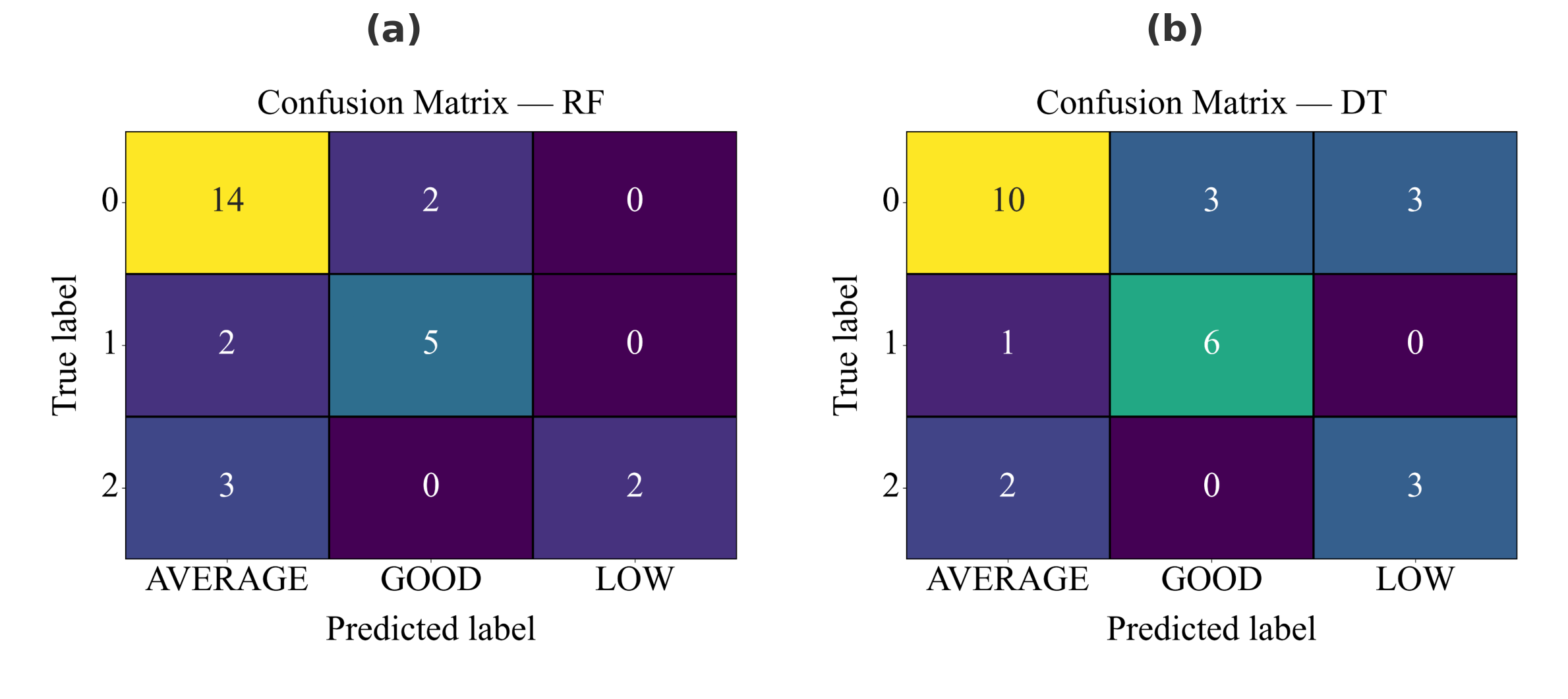} 
\caption{Confusion matrix (out-of-fold) for engineered $F_{\text{change}}$ features (a) and engineered $F_{\text{half}}$ features (b).}
\label{fig:cm-dt-rf}
\end{figure}

\subsection{Classification with $F_{\text{half}}$ Features}

\begin{figure}[ht!]
\centering
\includegraphics[width=1\linewidth]{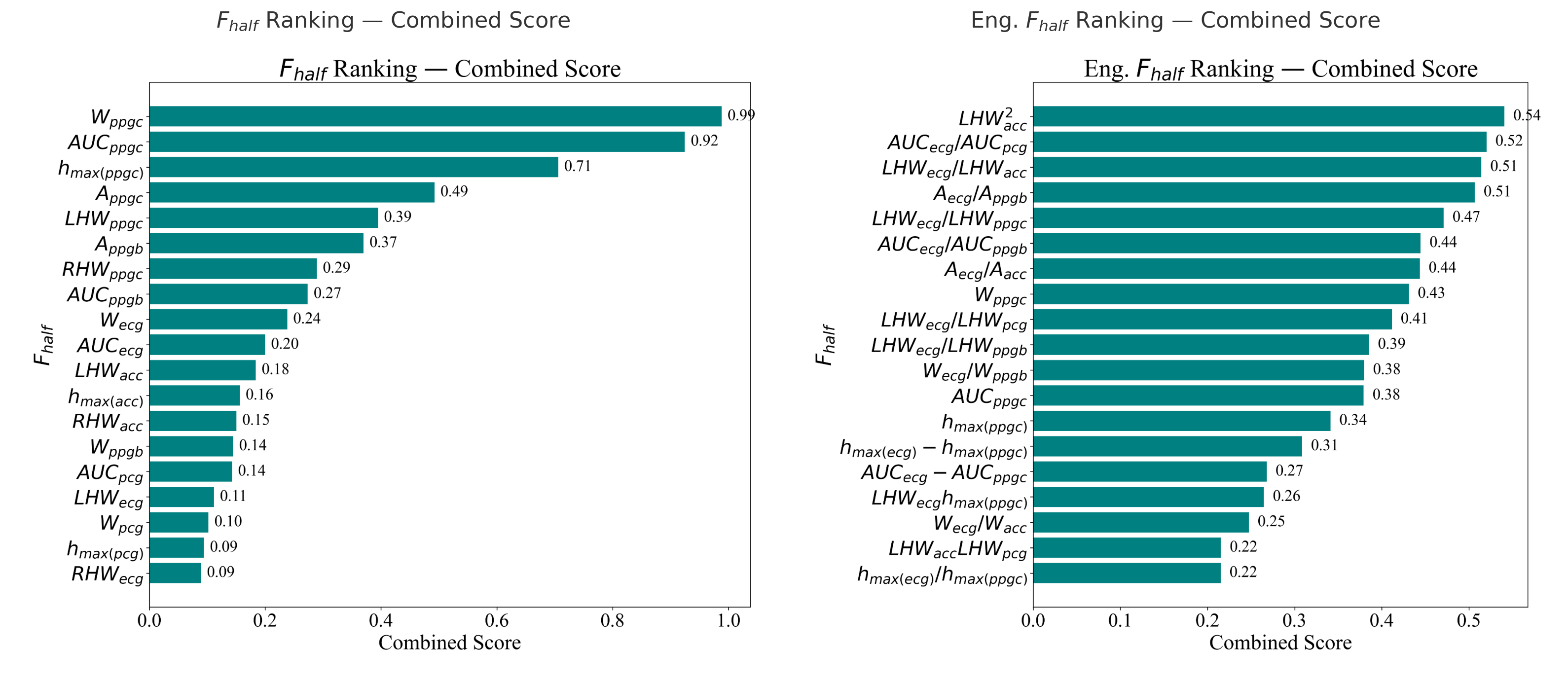}
\caption{Comparison of original vs engineered $F_{half}$ features ranked by Combined Score.}
\label{fig:feat-ranking-th}
\end{figure}

Figure~\ref{fig:feat-ranking-th} contrasts the top-20 original $F_{\text{half}}$ features with their engineered versions.  
Engineered descriptors involving left-half width and skewness across modalities consistently ranked higher,  
outperforming raw $F_{\text{half}}$ measures. As in the $F_{\text{change}}$ case, nonlinear and cross-modal combinations  appear to better capture temporal asymmetry of the waveforms, improving discrimination between classes.

\begin{table*}[htbp]
\centering
\resizebox{\textwidth}{!}{%
\begin{tabular}{l|ccc|ccc|ccc}
\hline
\multirow{2}{*}{Model} &
\multicolumn{3}{c|}{Accuracy} &
\multicolumn{3}{c|}{Macro-F1} &
\multicolumn{3}{c}{Balanced Accuracy} \\
 & S1 & S2 & S3 & S1 & S2 & S3 & S1 & S2 & S3 \\
\hline
LogReg   & 0.57$\pm$0.21 & 0.60$\pm$0.23 & 0.60$\pm$0.15 &
           0.51$\pm$0.25 & 0.52$\pm$0.29 & 0.56$\pm$0.17 &
           0.59$\pm$0.24 & 0.58$\pm$0.26 & 0.63$\pm$0.17 \\
SVM-RBF  & 0.48$\pm$0.18 & 0.68$\pm$0.06 & 0.49$\pm$0.18 &
           0.46$\pm$0.18 & 0.55$\pm$0.17 & 0.45$\pm$0.21 &
           0.53$\pm$0.24 & 0.62$\pm$0.18 & 0.56$\pm$0.17 \\
DT       & 0.57$\pm$0.10 & 0.43$\pm$0.21 & 0.67$\pm$0.17 &
           0.41$\pm$0.16 & 0.35$\pm$0.17 & 0.62$\pm$0.24 &
           0.49$\pm$0.19 & 0.37$\pm$0.18 & 0.68$\pm$0.23 \\
RF       & 0.57$\pm$0.10 & 0.57$\pm$0.06 & 0.75$\pm$0.14 &
           0.40$\pm$0.17 & 0.29$\pm$0.09 & 0.55$\pm$0.25 &
           0.49$\pm$0.19 & 0.38$\pm$0.10 & 0.63$\pm$0.22 \\
kNN      & 0.53$\pm$0.17 & 0.65$\pm$0.10 & 0.63$\pm$0.14 &
           0.31$\pm$0.19 & 0.43$\pm$0.16 & 0.53$\pm$0.18 &
           0.37$\pm$0.20 & 0.52$\pm$0.15 & 0.58$\pm$0.16 \\
\hline
\end{tabular}%
}
\caption{Model performance comparison under three $F_{half}$
-based feature selection strategies: global top five (S1), per-modality top five (S2), and engineered top three (S3). Results are expressed as mean $±$ SD using 5-fold cross-validation.}\label{tableII}
\end{table*}

Table~\ref{tableII} summarizes the 
classification results obtained using  $F_{\text{half}}$. Three complementary feature selection 
strategies were evaluated: (S1) selecting the five globally top-ranked features,  (S2) selecting the top-ranked feature per modality and (S3) selecting engineered features. 

Using the global top five features (Table~\ref{tableII} S1) produced 
moderate results overall. LogReg reached a macro-F1 of $0.51 \pm 0.25$ with balanced accuracy 
of $0.59 \pm 0.25$, while RF and DT achieved slightly lower values (macro-F1 $\approx 0.40$). 
kNN and SVM-RBF underperformed in this setting, suggesting that concentrating only on global 
ranking overemphasizes features from a single modality (here PPG\textsubscript{c}).

When selecting the top-ranked feature per modality (Table~\ref{tableII} S2), 
performance improved in terms of stability and multimodal balance. 
SVM-RBF performed best, with accuracy $0.68 \pm 0.07$ and macro-F1 $0.55 \pm 0.17$, 
followed by LogReg and kNN. Tree-based methods (DT, RF) lagged behind, 
likely due to sensitivity to noise in individual folds. 
This strategy ensured each biosignal contributed equally, and proved more robust 
than relying on globally top-ranked features alone.

Model comparison based on three engineered $F_{\text{half}}$ features is summarized in 
Table~\ref{tableII} S3. Decision Trees (DT) achieved the strongest 
macro-F1 performance ($0.62 \pm 0.25$), slightly ahead of Logistic Regression 
($0.56 \pm 0.17$) and Random Forests ($0.55 \pm 0.25$). $k$NN also reached 
reasonable values ($0.53 \pm 0.18$), while SVM-RBF underperformed with 
macro-F1 below $0.45$. These results show that tree-based methods benefited 
most from nonlinear feature engineering, whereas kernel methods were more 
sensitive to the small dataset size and the engineered feature space.

The out-of-fold confusion matrix for the best-performing model (DT) is shown 
in Fig.~\ref{fig:cm-dt-rf}. The \texttt{AVERAGE} class was classified most 
reliably (10/16 correct), whereas \texttt{GOOD} reached 6/7 correct, and 
\texttt{LOW} remained the most challenging (3/5 correct, with 2 misclassified 
as \texttt{AVERAGE}). This imbalance reflects the limited sensitivity of 
engineered $F_{\text{half}}$ features to low-recovery cases, but also demonstrates 
clear discriminative value for distinguishing average vs good recovery. 

Engineered features further allowed us to reduce the feature set to a compact, 
interpretable subset while still maintaining competitive performance. 
Because the dataset was relatively small, the differences between 
strategies were not very large. Still, different 
ways of choosing features emphasize different aspects of recovery dynamics, 
and the best choice depends on whether the priority is stronger prediction, 
balanced use of modalities, or simpler models.

\subsection{Comparison of $F_{\text{change}}$ and $F_{\text{half}}$ feature sets}

A direct comparison highlights complementary strengths of the two feature families.  
$F_{\text{change}}$ features summarize recovery as a net displacement between start and end points.  
Their main advantage is simplicity and interpretability: they indicate “how much” recovery occurred.  
However, their discriminative power is limited, with best macro-F1 values around 0.46 (RF) in the raw setting, rising to 0.62 when nonlinear engineering is applied.  

By contrast, $F_{\text{half}}$ features give the analysis at a standardized, physiologically meaningful time point.  
This leads to sharper between-class differences and stronger baseline performance:  
LogReg and SVM-RBF achieved macro-F1 values around 0.51–0.55, already exceeding raw $F_{\text{change}}$.  
When engineered $F_{\text{half}}$descriptors are used, performance remained competitive, with DT and RF reaching macro-F1 $\approx$0.55–0.62.  

The two sets thus emphasize different aspects of recovery dynamics.  
$F_{\text{change}}$ is sensitive to long-timescale adaptation, and benefits most from nonlinear transformations and cross-modality combinations.  
$F_{\text{half}}$ exploits temporal anchoring to expose short-to-intermediate scale differences between classes, and performs well with simpler linear models.  

Overall, $F_{\text{half}}$ features are more robust in small-sample settings and align better with physiologically interpretable markers of recovery, while engineered $F_{\text{change}}$ features offer additional discriminative power by capturing nonlinear, long-timescale changes.

\section{Discussion}

Multifractal measures have consistently revealed physiologically relevant structure in cardiovascular time series~\cite{Ivanov1999Multifractality,Goldberger1990Fractals}. In our study, post-exercise recovery emerges as a natural experiment to probe cardiovascular regulation: scaling properties of ECG, PPG, SCG, and PCG signals display transient shifts that gradually return toward baseline. 
These shifts occur with different time constants across electrical (ECG), mechanical (SCG), optical–mechanical (PPG), and acoustic–mechanical (PCG) subsystems, highlighting recovery as a coordinated re-synchronization of coupled physiological processes rather than a uniform return to baseline. Although PPG is often associated with vascular responses, it is fundamentally an optical–mechanical signal that reflects pulsatile changes in blood volume due to cardiac ejection.

Our results demonstrate that even with a compact set of features, non-trivial three-class discrimination of recovery states can be achieved. Features drawn directly from individual modalities retain clear physiological meaning—for example, ECG-derived $F_{\text{change}}$ captures electrical control, while PPG asymmetry reflects vascular compliance. Tree-based classifiers improved markedly when provided with nonlinear interaction terms and cross-modality combinations, suggesting that recovery dynamics involve nonlinear couplings between vascular and mechanical responses. This improvement is especially important for identifying the LOW recovery group—clinically the most relevant category—because impaired recovery often precedes overt pathology~\cite{cole1999heart}.

Compared with previous recovery studies that focused mainly on HRV or ECG descriptors%
~\cite{Clifford2012,Acharya2017}, our approach integrates optical (PPG), mechanical 
(SCG), and acoustic (PCG) modalities in addition to ECG. This multimodal perspective 
supports the view that fitness and resilience are system--level properties, emerging 
from the interplay of diverse physiological subsystems. These findings align with 
broader evidence from complex systems and physiology, where integrating complementary 
signals exposes dependencies that single--channel analyses may miss%
~\cite{Stanley1999Scaling,Goldberger1990Fractals}.

Despite promising performance, several limitations temper our conclusions. The dataset size ($N=28$) limits statistical power and risks overfitting, especially when nonlinear interactions expand the feature space. Stratified cross-validation and class-weight balancing helped stabilize estimates but do not substitute for validation on an independent cohort. Motion artifacts in SCG and PCG remain a challenge for early recovery phases, and further signal-quality checks may improve robustness. Engineered features, while interpretable, cannot fully capture latent patterns that modern representation-learning methods might reveal.

From a diagnostic and health-monitoring perspective, these findings suggest two complementary strategies. For clinical interpretation or wearable devices where transparency is critical, modality-specific features may support trust and insight. For automated triage or fitness monitoring, engineered features capturing nonlinear couplings may maximize performance. Future research should combine physiologically grounded descriptors (e.g., $F_{\text{half}}$, HRR) with engineered features, expand datasets to diverse populations, and test real-time analysis pipelines. Ultimately, integrating multifractal dynamics with multimodal sensing could provide early warning signals of cardiovascular dysregulation before overt symptoms appear—a preparatory step toward advanced diagnostic frameworks that link signal complexity to human health.

\section{Conclusion}

We studied recovery dynamics after exercise using multifractal features 
from multimodal cardiovascular signals (ECG, PPG, SCG, PCG). Two types of 
temporal-change features were compared: $F_{\text{change}}$, which captures 
long-term adaptation between pre- and post-exercise states, and 
$F_{\text{half}}$, which measures recovery at a standardized half-recovery 
time. Both families provided useful but distinct information.
$F_{\text{half}}$ features gave robust and interpretable discrimination in their raw form and improved further with engineered terms.
$F_{\text{change}}$ features were weaker in raw form but gained strongly from nonlinear and cross-modality interactions, where ensemble methods such as Random Forests performed best.

Our results show that meaningful three-class discrimination of recovery state 
is possible with a small set of features, even in a limited dataset. The 
analysis highlights that recovery is a coupled process across multiple 
physiological subsystems,and that different feature strategies show different aspects:
$F_{\text{half}}$ captures recovery at a fixed reference point, while
$F_{\text{change}}$ describes longer-term changes that become clearer when nonlinear interactions are included.

Future work should confirm these findings in larger and more diverse groups, test robustness under real-world conditions, and develop hybrid approaches that combine physiologically based and engineered features. This could lead to reliable tools for monitoring recovery in clinical and sports settings and help with early detection of abnormal adaptation

\section*{Acknowledgements}

This research is supported by the Science Fund of the Republic of Serbia, Grant No.~7754338, Multi-SENSor SysteM and ARTificial intelligence in service of heart failure diagnosis -- SensSmart. We acknowledge the support from the Ministry of Science, Technological Development and Innovation of the Republic of Serbia, Grant No.~451-03-136/2025-03/200017.

\appendix

\section{Determination of Half–Recovery Features}

The preparatory phase for classification consists of feature definition, selection, and relevance ranking (Fig.~\ref{fig:workflow}). 

Recovery dynamics of each multifractal feature $F(t)$ were modeled using several 
candidate functions, including single-exponential decay (E1), exponential with offset 
(E1C), double-exponential with offset (E2C), and polynomial regression when 
exponential models were inadequate. The best model was selected for each subject 
and feature based on goodness-of-fit criteria \cite{cole1999heart,Shaffer2017HRV}. In exponential models, the fitted 
parameter $\tau$ represents the relaxation time constant, characterizing how quickly 
the feature approaches its equilibrium state. In some cases, however, no stable $\tau$ 
could be obtained, particularly for polynomial fits or poorly defined trajectories.

To provide a consistent benchmark across all subjects, we therefore defined for each 
feature and modality a cohort-level median half-recovery time, 
$T_{1/2}^{\mathrm{cohort}}$, calculated from subjects with finite and reliable 
time constants. The fitted curve $F(t)$ for each subject was then evaluated at this 
reference time 
$F_{\text{half}} = F(T_{1/2}^{\mathrm{cohort}})$. The absolute value of $F_{T_{1/2}}$ reflects the recovery state at a physiologically meaningful time.

Anchoring all measurements to the cohort-defined $F_{\text{half}}$ avoided inconsistencies 
caused by missing or divergent individual time constants, and ensured that descriptors 
were directly comparable across subjects and modalities. This approach provided a 
robust basis for subsequent feature ranking and classification analyses.

Physiologically, the relaxation time $\tau$ indicates how rapidly the system 
returns to equilibrium, with larger values corresponding to slower recovery. 
The half-time $T_{1/2}^{\mathrm{cohort}}$ serves as a standardized reference 
point across individuals, even when subject-specific estimates are unreliable. 
The absolute value $F_{\text{half}}$ characterizes the recovery state itself, 
while the gap $\Delta F_{\text{half}}$ reflects the progress achieved by that 
time relative to the immediate post-exercise condition. Together, these descriptors 
combine robustness with interpretability, enabling meaningful comparison of recovery 
dynamics across features, modalities, and subjects.

\section{Per-fold Model Performance for Engineered $F_{\text{change}}$ Features}
\label{app:folds-dp}

For completeness, Tables~\ref{tab:dp-fold-SCG}, \ref{tab:dp-fold-f1}, and \ref{tab:dp-fold-balSCG}
report the fold-wise results (5-fold stratified CV) for each model. These complement the
averaged values presented in the main text.

Overall, the fold-wise results confirm the trends observed in the mean performance:
\begin{itemize}
    \item RF achieves the strongest overall performance, with multiple folds reaching near-perfect accuracy, but also shows higher variance across folds.
    \item DT benefits from engineered nonlinear features but is less consistent, with notable fluctuations between folds.
    \item LogReg is comparatively stable across folds, providing moderate accuracy and macro-F1.
    \item SVM-RBF is inconsistent, with strong performance in some folds but weak in others.
    \item kNN performs well when class separation is clearer, but degrades in folds dominated by boundary cases.
\end{itemize}

These observations highlight that engineered $F_{\text{change}}$ features particularly
benefit ensemble/tree-based methods (e.g., RF), which can leverage nonlinear interactions
and cross-modality products, while linear models remain more stable but less expressive.

\begin{table}[htbp]
\centering
\caption{Per-fold accuracy for each model (engineered $F_{\text{change}}$ features).}
\label{tab:dp-fold-SCG}
\begin{tabular}{lccccc}
\hline
Model & Fold 1 & Fold 2 & Fold 3 & Fold 4 & Fold 5 \\
\hline
DT       & 0.50 & 0.67 & 1.00 & 0.60 & 0.60 \\
LogReg   & 0.50 & 0.67 & 0.83 & 0.40 & 0.60 \\
RF       & 0.67 & 0.67 & 1.00 & 0.60 & 0.80 \\
SVM-RBF  & 0.50 & 0.83 & 0.33 & 0.40 & 0.40 \\
kNN      & 0.83 & 0.67 & 0.67 & 0.60 & 0.40 \\
\hline
\end{tabular}
\end{table}

\begin{table}[htbp]
\centering
\caption{Per-fold macro-F1 for each model (engineered $F_{\text{change}}$ features).}
\label{tab:dp-fold-f1}
\begin{tabular}{lccccc}
\hline
Model & Fold 1 & Fold 2 & Fold 3 & Fold 4 & Fold 5 \\
\hline
DT       & 0.52 & 0.72 & 1.00 & 0.25 & 0.61 \\
LogReg   & 0.52 & 0.66 & 0.82 & 0.30 & 0.49 \\
RF       & 0.47 & 0.49 & 1.00 & 0.25 & 0.56 \\
SVM-RBF  & 0.36 & 0.87 & 0.33 & 0.30 & 0.39 \\
kNN      & 0.63 & 0.72 & 0.67 & 0.25 & 0.39 \\
\hline
\end{tabular}
\end{table}

\begin{table}[htbp]
\centering
\caption{Per-fold balanced accuracy for each model (engineered $F_{\text{change}}$ features).}
\label{tab:dp-fold-balSCG}
\begin{tabular}{lccccc}
\hline
Model & Fold 1 & Fold 2 & Fold 3 & Fold 4 & Fold 5 \\
\hline
DT       & 0.50 & 0.78 & 1.00 & 0.33 & 0.78 \\
LogReg   & 0.50 & 0.78 & 0.89 & 0.44 & 0.56 \\
RF       & 0.58 & 0.56 & 1.00 & 0.33 & 0.67 \\
SVM-RBF  & 0.50 & 0.89 & 0.50 & 0.44 & 0.44 \\
kNN      & 0.67 & 0.72 & 0.72 & 0.33 & 0.44 \\
\hline
\end{tabular}
\end{table}

\section{Per-fold Model Performance for $F_{\text{half}}$ Features}
\label{app:folds}

In Tables~\ref{tab:fold-SCG}, \ref{tab:fold-f1}, and \ref{tab:fold-balSCG}
the fold-wise results (5-fold stratified CV) for $F_{\text{half}}$ engineered features for each classification model are presented. These results complement the averaged values presented in the main text.

The the fold-wise results confirm the trends observed in the mean performance:
\begin{itemize}
    \item DT and RF showed the highest variance across folds, occasionally reaching perfect classification but also dropping substantially in other folds. This reflects their sensitivity to the small dataset size.
    \item LogReg was more stable, maintaining moderate accuracy and macro-F1 across all folds, which explains its strong averaged performance in the main text.
    \item SVM-RBF achieved strong results in certain folds (notably Fold 2), but was inconsistent overall.
    \item kNN tended to perform well when class clusters were clear (Fold 1–2), but underperformed when boundary cases dominated (Fold 5).
\end{itemize}

These observations underline that while ensemble and tree-based methods can capture nonlinear
patterns, their performance is less stable under cross-validation in small-sample settings.
Linear models (LogReg) provided more robust and interpretable outcomes across folds.

\begin{table}[htbp]
\centering
\caption{Per-fold accuracy for each model.}
\label{tab:fold-SCG}
\begin{tabular}{lccccc}
\hline
Model & Fold 1 & Fold 2 & Fold 3 & Fold 4 & Fold 5 \\
\hline
DT       & 0.50 & 0.67 & 1.00 & 0.60 & 0.60 \\
LogReg   & 0.50 & 0.67 & 0.83 & 0.40 & 0.60 \\
RF       & 0.67 & 0.67 & 1.00 & 0.60 & 0.80 \\
SVM-RBF  & 0.50 & 0.83 & 0.33 & 0.40 & 0.40 \\
kNN      & 0.83 & 0.67 & 0.67 & 0.60 & 0.40 \\
\hline
\end{tabular}
\end{table}

\begin{table}[htbp]
\centering
\caption{Per-fold macro-F1 for each model.}
\label{tab:fold-f1}
\begin{tabular}{lccccc}
\hline
Model & Fold 1 & Fold 2 & Fold 3 & Fold 4 & Fold 5 \\
\hline
DT       & 0.52 & 0.72 & 1.00 & 0.25 & 0.61 \\
LogReg   & 0.52 & 0.66 & 0.82 & 0.30 & 0.49 \\
RF       & 0.47 & 0.49 & 1.00 & 0.25 & 0.56 \\
SVM-RBF  & 0.36 & 0.87 & 0.33 & 0.30 & 0.39 \\
kNN      & 0.63 & 0.72 & 0.67 & 0.25 & 0.39 \\
\hline
\end{tabular}
\end{table}

\begin{table}[htbp]
\centering
\caption{Per-fold balanced accuracy for each model.}
\label{tab:fold-balSCG}
\begin{tabular}{lccccc}
\hline
Model & Fold 1 & Fold 2 & Fold 3 & Fold 4 & Fold 5 \\
\hline
DT       & 0.50 & 0.78 & 1.00 & 0.33 & 0.78 \\
LogReg   & 0.50 & 0.78 & 0.89 & 0.44 & 0.56 \\
RF       & 0.58 & 0.56 & 1.00 & 0.33 & 0.67 \\
SVM-RBF  & 0.50 & 0.89 & 0.50 & 0.44 & 0.44 \\
kNN      & 0.67 & 0.72 & 0.72 & 0.33 & 0.44 \\
\hline
\end{tabular}
\end{table}

\nocite{*}
\bibliographystyle{aipnum4-1}
\bibliography{CHAOSMFNL} 

\end{document}